\documentclass[%
 aip,
 amsmath,amssymb,
 reprint,%
]{revtex4-1}

\usepackage{graphicx}
\usepackage{dcolumn}
\usepackage{bm}

\usepackage[utf8]{inputenc}
\usepackage[T1]{fontenc}
\usepackage{mathptmx}
\usepackage{etoolbox}
\usepackage{tikz}

\makeatletter
\def\@email#1#2{%
 \endgroup
 \patchcmd{\titleblock@produce}
  {\frontmatter@RRAPformat}
  {\frontmatter@RRAPformat{\produce@RRAP{*#1\href{mailto:#2}{#2}}}\frontmatter@RRAPformat}
  {}{}
}%
\makeatother
\begin{document}

\preprint{AIP/123-QED}

\title{Precision Evaluation Criteria for Simulation Algorithms in Infinite Systems: A Network Model-Based Approach}
\author{Yonglong Ding}
\affiliation{School of Physics, Beijing Institute of Technology, Beijing 100081, China}
\affiliation{Beijing Computational Science Research Center, Beijing 100193, China}
 \email{ylding@csrc.ac.cn.}

\date{\today}

\begin{abstract}
As the particle count escalates, the computational demands of diverse simulation algorithms surge, paralleled by a marked enhancement in accuracy. The question arises whether this heightened precision asymptotically dwindles towards zero or plateaus at a finite constant. To address this, this work introduces an approach that translates infinite systems into finite-node network architectures, providing a rigorous framework for assessing this question.
Employing the Monte Carlo algorithm's application to the Ising model as a case study, this paper demonstrate that despite the simulation's extension to an infinite lattice size, a fundamental error bound persists. This work explicitly derive this lower bound on the error, offering a quantitative understanding of the algorithm's limitations in the limit of infinite scale.
Furthermore, I extend this methodology to Molecular Dynamics simulations, exemplified through its application to battery systems. This conversion strategy not only underscores the generality of this approach but also highlights its practical significance in guiding the optimization of simulation algorithms. Moreover, it offers insights into estimating micro-level information from macro-level data. The crucial information of Molecular Simulation, namely the potential energy, has been quickly estimated.
\end{abstract}

\maketitle


\section{Introduction}
Drawing parallels with the seminal principles of Fourier transform, Ding\cite{Ding202401,Ding202402} has formulated an algorithm that elegantly maps an infinite lattice model onto a finite-node network architecture. This transformation meticulously reduces the expanse of the infinite lattice to a tractable, finite ensemble of interconnected network nodes, while concurrently transforming the rudimentary interactions among lattice points within the model into intricate, multifaceted interactions between the nodes of the resultant network structure.
To elucidate the intricacies of these complex interactions, Ding employs Monte Carlo algorithms, harnessing their powerful statistical sampling capabilities to rigorously characterize and quantify the dynamics at play. This methodological refinement has culminated in noteworthy accomplishments within the realm of fractal geometry and phase transition research.

The Monte Carlo algorithm\cite{RevModPhys.73.33,sarrut2021advanced,luo2022hybrid} stands as a highly efficient tool, evolving into tailored methodologies such as projector quantum Monte Carlo\cite{schwarz2017projector} Continuous-time quantum Monte Carlo\cite{rubtsov2005continuous} diagrammatic Monte Carlo\cite{prokof2008fermi} and determinant Monte Carlo\cite{moutenet2018determinant}, tailored to specific problems, thereby occupying a pivotal position in scientific research endeavors. Similarly, Molecular Dynamics\cite{brooks2021classical,van1998validation,peter2004estimating} simulations exhibit remarkable efficiency, empowering researchers to delve into fundamental laws and principles, with extensive applications spanning chemistry\cite{park2004calculating} and biology\cite{csenol2024novel}. Nevertheless, both these algorithms confront a formidable challenge: the computational demands escalate dramatically with the increase in system size\cite{ortega2015fpga,yang2019high,preis2009gpu,meredith2009accuracy}, posing limitations on our ability to gain nuanced insights into the behavior of infinite systems. 

Sec.~\ref{sec:1} comprehensively delineate the overarching methodology, initiating with a meticulous exposition on the transformative process whereby myriad particles are systematically converted into a discrete set of network nodes. This conversion is predicated on the intricate interconversion dynamics between particles of diverse categories, which in turn inform the direct transformational relationships established between the resultant network nodes. Furthermore, this paper devise an innovative approach to link these nodes with varying bonds, where each bond represents a specific interaction pattern, thereby reflecting the underlying complexity of the system.
To substantiate the rigor and efficacy of simulation algorithm, this work derive a robust lower error bound, leveraging specialized network structures. This lower bound serves as a crucial benchmark for evaluating the algorithm's performance and ensuring its reliability across diverse applications.
Sec~\ref{sec:2a} delve into the specific nuances of the transformation process for the Ising model, demonstrating the algorithm's exceptional efficiency in precisely determining the lower error bound of Monte Carlo simulations within this complex statistical physics framework. This achievement underscores the algorithm's potential to significantly advance the accuracy and precision of simulation-based research in this domain.
Sec~\ref{sec:2b} commence by presenting the intricate details of the transformation process, ensuring a comprehensive understanding of the underlying mechanics. Subsequently, this paper provide a more elaborate illustration, utilizing the exemplary case of a battery system. This case study not only showcases the versatility and applicability of this algorithm but also highlights its potential to yield novel insights into the dynamic behavior of complex molecular systems. Furthermore, it offers insights into estimating micro-level information from macro-level data.

\section{Algorithm}
\label{sec:1}

Sec.~\ref{sec:1} comprehensively summarizes the overall handling and operational procedures of the algorithm. To delve into the properties of systems encompassing an infinite number of particles, Ding\cite{Ding202401,Ding202402} devised an innovative strategy that involves categorizing all particles or lattice sites within the infinite system in a systematic and principled manner. This transformation reduces the complexity of an infinite system to a finite set of well-defined categories, facilitating analysis. In scenarios where particle counts are immense and inherently stochastic, such a categorization framework emerges as both rational and practical. Critically, it circumvents the imposition of rigid boundaries, thereby obviating the need for periodic boundary conditions, which can introduce artifacts into the analysis.
This approach echoes the simplification strategies employed in chemistry when constructing reaction equations, yet it transcends those conventions by offering a heightened degree of flexibility. The methodology presented herein is tailored to accommodate the unique requirements of each problem, enabling the adoption of diverse classification schemes. For instance, in the chemical reaction $A+B\to C+D$, this approach would not be constrained to the traditional binary categorization but would instead explore the possibility of segmenting $A$ and $B$ into three distinct categories: $A$, $B$, and a hypothetical intermediate or complex state $AB$, reflecting a deeper understanding of the underlying reaction mechanisms.

Upon implementing this categorization method, an infinite system is reduced to a manageable finite collection of categories. However, the subsequent interactions between these various categories frequently exhibit a profound complexity. To delve into and comprehend this intricate network of interactions, this algorithm adopt a network-based approach, wherein each category is transformed into a distinct node within a network. Nodes that represent categories with interactional or transformational relationships are then directly interconnected, forming a clear and comprehensive network architecture. 

Accurately calculating this network structure and achieving results that mirror or closely approximate those of an infinite system poses a formidable challenge. To address this intricacy, this approach commences from the nodes themselves, harnessing the principles of detailed balance and network theory. In particular, this approach employ Monte Carlo algorithms to meticulously examine the relationships among distinct nodes. Notably, in many instances, the connections between network nodes are not isolated but intertwined; the relationships a given node shares with its various interconnected counterparts are mutually dependent. By meticulously probing each node individually, we can comprehensively unravel the complex network of interactions that characterizes the entire network. Overall, this work aim to delve into the relationships between network nodes through the lens of detailed balance.

Upon meticulously constructing the intricate network of interactions that permeates the entire network, its complex interrelationships unravel a plethora of properties, highlighting the presence of unique network structures. Through a rigorous analysis and computation of these structures, this paper delves into the properties inherent in infinite systems. Returning to the core of this discussion, I introduce a criterion via this algorithm: Assuming that various simulation methodologies are adept at computing systems of boundless dimensions or with an infinite count of particles, this algorithm determine whether the computational accuracy converges to zero or asymptotically approaches a constant value.

\section{Calculation}
\label{sec:2}
The Sec.~\ref{sec:1} provided a concise overview of the algorithm's conceptual framework. Subsequently, this work will delve deeper into the specifics of the algorithm through illustrative examples, thereby validating its feasibility. Specifically, Sec.~\ref{sec:2a} focuses on verifying the Monte Carlo algorithm, while Sec.~\ref{sec:2b} addresses the validation of the molecular dynamics algorithm.

\subsection{Monte Carlo}
\label{sec:2a}

In this section, this work will delve deeper into the specific implementation of the Monte Carlo algorithm, exemplified by its application in calculating the Ising model. Furthermore, this work will undertake a structural analysis of the network to elucidate a potential lower bound on the error achievable through the Monte Carlo algorithm.

The two-dimensional Ising model constitutes a robust stochastic process framework, intricately crafted to portray the intricate interplay between spins situated on a two-dimensional lattice and the consequential phase transitions that ensue. Fundamentally, it posits that each lattice site harbors a spin, constrained to binary states: +1 signifying spin-up and -1 denoting spin-down. Typically, this model embraces the square lattice as its underlying geometric archetype, yet its versatility extends to accommodate a variety of two-dimensional lattice configurations. Within the square lattice's context, each node is intimately tied to its four immediately adjacent neighbors—above, below, to the left, and to the right—via the mediating force of interaction energy.

The Ising model assumes that only interactions occur between nearest-neighbor spins, which can be either ferromagnetic (favoring parallel alignment of adjacent spins) or antiferromagnetic (favoring antiparallel alignment). The interaction energy is typically denoted as $-Js_{i}s_{j}$, where $J$ represents the strength of the interaction ($J > 0$ for ferromagnetic, $J < 0$ for antiferromagnetic), and $s_{i}$ and $s_{j}$ are the spin values of the two adjacent sites.
The total energy, or Hamiltonian, of the system encapsulates the cumulative effect of these interactions between neighboring spin pairs. It is succinctly expressed as 
$H = -J \sum < i,j> s_{i} s_{j}$, where the summation $\sum<i,j>$ elegantly iterates over all such pairs, providing a comprehensive account of the system's energetic landscape.

When the system's temperature dips below a critical threshold, denoted as $T_{c}$, the two-dimensional Ising model undergoes a pivotal transition from a state of disorder to one of order.
In this ordered configuration, the spins coalesce into a coherent arrangement, either all pointing upwards or downwards, giving rise to a discernible macroscopic magnetic property.

\begin{figure}
\centering
\begin{tikzpicture}

\scope[nodes={inner sep=0,outer sep=0}]
\node[anchor= east] (a)
  {\includegraphics[width=4cm]{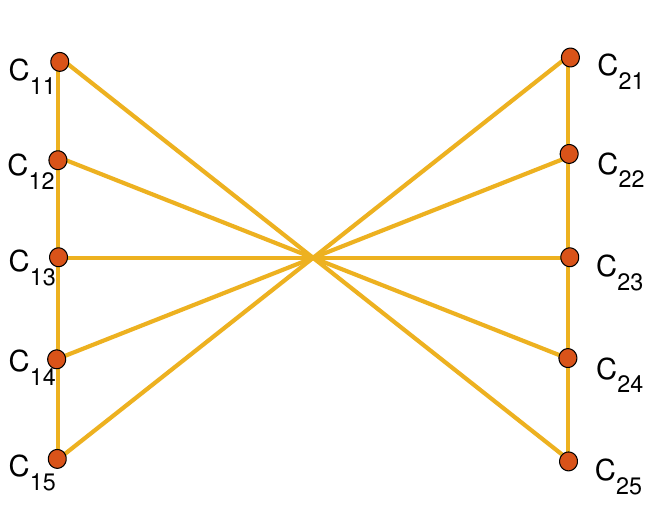}};
\node[anchor= west] (b)
  {\includegraphics[width=4cm]{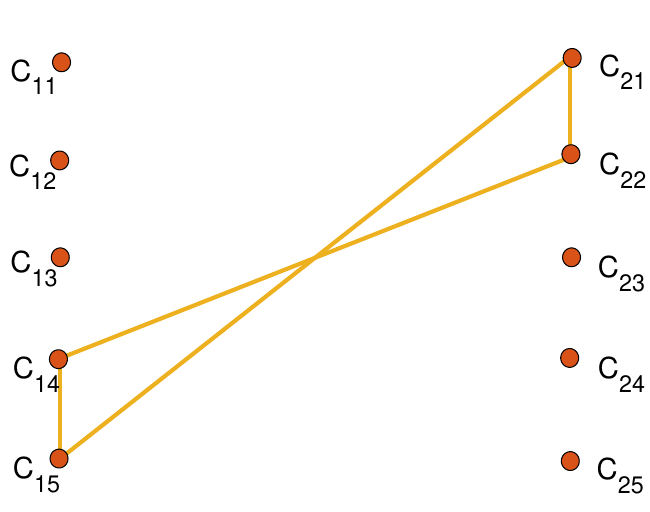}};
\endscope
\foreach \n in {a,b} {
  \node[anchor=west] at (\n.north west) {(\n)};
}

\end{tikzpicture}

\caption{\label{fig:1} The red dots symbolize the network nodes, each corresponding to a distinct type of lattice point within the two-dimensional Ising model. Meanwhile, the yellow bonds illustrate the transformational relationships among these various nodes. (a) (b) represents two distinct and stable network configurations observed within the Ising model. $C$ represents the weight of the node. (a) signifies the direct transformation relationship that holds true for all potential nodes. Meanwhile, (b) encompasses the two nodes with the lowest energy and their corresponding counterparts adhering to the principle of detailed balance. Evidently, the magnetic induction intensity observed in reality falls below that calculated by this particular network configuration. }
	\label{fig2}
 
\end{figure}

For each lattice point, it is surrounded by exactly four mutually interacting nearest-neighbor lattice points, and the lattice point itself can have a spin that is either upward or downward. Based on these criteria, the lattice points are categorized into 10 distinct groups. For any individual lattice point, it falls into one of 5 subcategories based on the number of its nearest-neighbor lattice points that share the same spin. When also considering the inherent spin property of each lattice point, the total number of categories expands to 10. These 10 categories are then mapped onto 10 corresponding network nodes, with connections established between nodes that can be directly transformed into each other. In this manner, a comprehensive network structure is formed. The 10 categories are denoted by $C_{mn}$, where $m$ signifies the type of spin, and $n$ represents the subcategorization based on the differing spins, with increasing numbers corresponding to decreasing energy levels within the respective categories, as shown in Fig.~\ref{fig:1} (a).

As depicted in the Fig~\ref{fig:1} (b), adhering to the principle of detailed balance, the nodes between the two layers of the network maintain a state of detailed balance. This enables us to derive a corresponding relationship between the weights of these two layers.
Turning our attention to the Monte Carlo algorithm, when a flip is executed, a randomly selected lattice point may undergo a transition. If such a transition occurs, this lattice point then becomes a node within the alternate layer. Concurrently, the nodes associated with the four neighboring lattice points surrounding this one also undergo a change. Consequently, a successful flip influences the alteration of network nodes associated with five lattice points in total.

Now, let's delve into a scenario where, under the premise of detailed balance, alterations in magnetic induction intensity are of interest. Traditional Monte Carlo algorithms, utilizing periodic boundary conditions, simulate finite-sized systems yet can still anticipate phase transitions. The question arises: how would the results pan out if the simulated system were of infinite extent?
This paper explores a network architecture that adheres to the principle of detailed balance, postulating that the magnetic induction intensity computed through actual Monte Carlo simulations would be no greater than that derived from this specialized network structure. This approach serves as a means to establish a lower bound for the inherent error.
This unique network structure commences with all spins aligned upwards at an initial, exceedingly low temperature. As the temperature gradually increases, select lattice points undergo spin flips, thereby giving rise to four distinct classes of lattice points, as vividly depicted in the accompanying Fig~\ref{fig:1} (b).

Certainly, the magnetic induction intensity in this hypothetical scenario is greater than that observed in actual Monte Carlo runs. The weight correspondence between the two-layer nodes, as dictated by detailed balance, with the focus solely on these four nodes, presents a valid lower bound.

Now, let's delve into the computation of this unique structure.

Firstly, we ensure particle conservation:
\begin{equation}
    C_{14}+ C_{15}+ C_{21}+ C_{22}=1
\end{equation}

Adhering to the principle of detailed balance, we derive:
\begin{equation}
   C_{21}= C_{15} e^{-8/T},
   C_{22}= C_{14} e^{-4/T} 
\end{equation}

To account for the spins of neighboring lattice points, this paper note:
The total number of neighboring lattice points with spins up is given by:
\begin{equation}
    4C_{15}+3C_{14}+ C_{22}=2-2k
\end{equation}

Here, $k$ signifies the proportionality factor for the negative bonds. A negative bond indicates that the two spins it connects are opposite, whereas a positive bond signifies that the spins it joins are  identical. And the total number of neighboring lattice points with spins down is:
\begin{equation}
    C_{14}+4 C_{21}+ 3C_{22}=2k
\end{equation}

By solving these equations in tandem, we can ascertain the values of the variables.

\begin{equation}
    \frac{C_{14}(1-3e^{-4/T})}{4}=\frac{k-C_{14}(1+e^{-4/T})}{1+e^{-8/T}}e^{-8/T}
\end{equation}

\begin{figure}
\centering

  {\includegraphics[width=8cm]{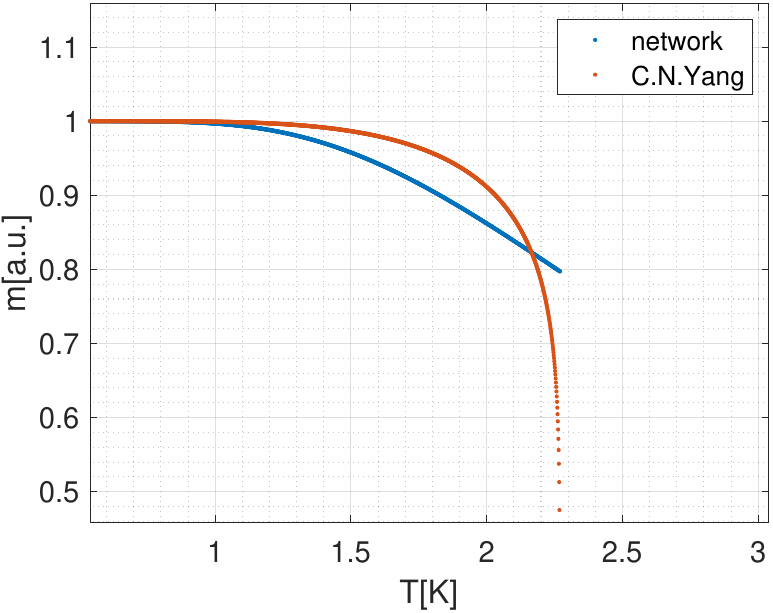}};

\caption{\label{fig:2}  The Temperature-Dependent Magnetic Susceptibility in the Two-Dimensional Ising Model. The red dots represent the analytical solution of the two-dimensional Ising model, elucidated by C.N.Yang\cite{yang1952spontaneous}. The blue line delineates an upper boundary for Monte Carlo simulations, as calculated by this work. Assuming the Monte Carlo algorithm computes this problem without any errors, the red line ought to lie beneath the blue line. }
	\label{fig2}
 
\end{figure}

As the two-dimensional Ising model boasts analytical solutions\cite{yang1952spontaneous}, we can straightforwardly contrast the computational outcomes with these solutions,as shown in Fig.~\ref{fig:2}. The blue line depicts the simulation outcomes of Monte Carlo within an infinite lattice model, whereas the red line corresponds to the analytical results. It is evident that there exists a discernible degree of error. This approach not only furnishes us with a lower bound for the errors inherent in the Monte Carlo algorithm's simulation of the two-dimensional Ising model but also illuminates avenues for algorithm optimization. For example, incorporating the flipping of multiple lattice points into the algorithm holds promise for enhancing its efficiency and accuracy. 

\subsection{Molecular Dynamics}
\label{sec:2b}
This section presents the outcomes of Molecular Dynamics, exemplified by the two-dimensional scenario, with the methodology readily applicable to multi-dimensional cases.
In a two-dimensional system, a particle's velocity is dissected into two orthogonal components, along the $x$ and $y$ axes. A value of 1 along any axis signifies motion in the positive direction, whereas -1 indicates motion in the opposite direction. This holds true for both axes, categorizing all possible particle states into 4 distinct scenarios: $C_{1}$ (1,-1), $C_{2}$ (1,1), $C_{3}$ (-1,1), and $C_{4}$ (-1,-1). These scenarios are then transformed into individual network nodes, as shown in Fig.~\ref{fig:2}(a).
When it comes to the evolution of particles within their respective nodes, the particle's acceleration direction follows a similar decomposition along the $x$ and $y$ axes. Positive acceleration along an axis is denoted by 1, and negative acceleration by -1. As a result, the effects imparted on particles can also be neatly partitioned into 4 categories: $C_{1}$ (1,-1), $C_{2}$ (1,1), $C_{3}$ (-1,1), and $C_{4}$ (-1,-1). Unlike particle velocities, however, the varying acceleration directions are embodied as bonds connecting the network nodes, each bond possessing a definitive orientation that signifies the probable transition path of the particle.

\begin{figure}
\centering
\begin{tikzpicture}

\scope[nodes={inner sep=1.5,outer sep=1.5}]
\node[anchor= east] (a)
  {\includegraphics[width=4cm]{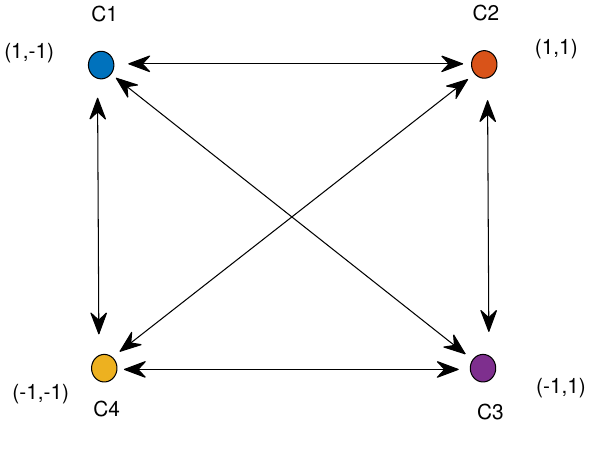}};
\node[anchor= west] (b)
  {\includegraphics[width=4cm]{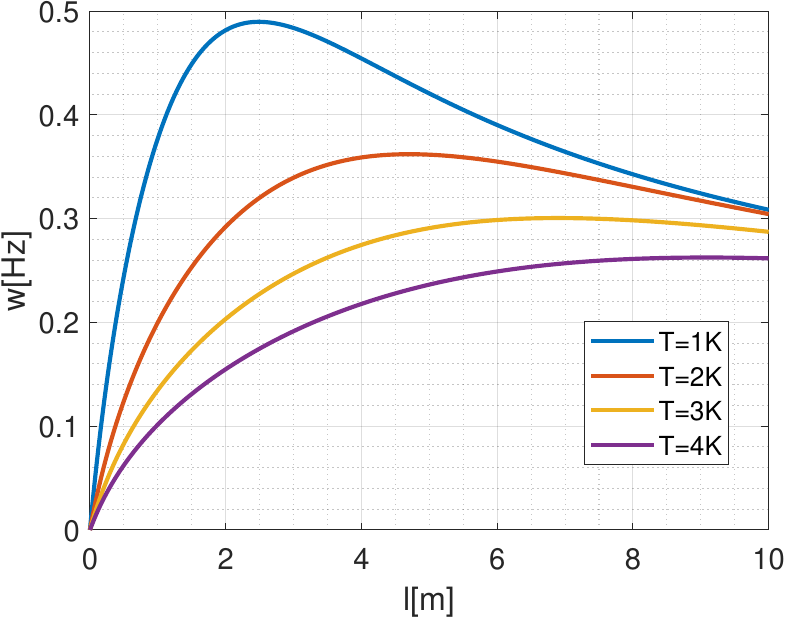}};
\endscope
\foreach \n in {a,b} {
  \node[anchor=west] at (\n.north west) {(\n)};
}

\end{tikzpicture}

\caption{\label{fig:3} Molecular Dynamics (a) The colored dots signify the network nodes derived from the transformation of all particles, with double arrows illustrating the interrelationships between 4 such nodes.
(b) This paper adopts the harmonic oscillator approximation to represent all electrons with the potential to migrate. By setting non-variable factors like particle mass to unity, this work obtain a graph that illustrates the trend of frequency variation with respect to distance within the battery. The distinct colors are utilized to signify varying temperatures. }
	\label{fig2}
 
\end{figure}

Hence, the direction of a particle's velocity does not necessarily alter under acceleration. If it does, the transition occurs along a distinct bond to a different network node. The 4 nodes are interconnected by all possible bonds.
At equilibrium, detailed balance prevails between the nodes, governed by the intricate interplay of intermolecular forces as described by Molecular Dynamics. This interplay facilitates the quantification of node weights. Specifically, for any particle residing within a node, we can ascertain the probabilities of its interactions with the 4 distinct bonds and the subsequent transitions to other nodes.
By juxtaposing these theoretically derived weights against experimentally measured ones, we can establish a lower bound on the precision of Molecular Dynamics simulations. Experimenting with various computational methodologies or network designs can yield diverse lower bounds. Notably, a higher lower bound signifies a closer alignment with the simulation outcomes of Molecular Dynamics in the limit of an infinite particle count.

Once transformed into a network architecture, direct positional details and velocity magnitudes are omitted. Nevertheless, the densities of particles across various directions can be indirectly inferred from the weights assigned to different nodes and the transition probabilities represented by the connecting bonds. Moreover, within a given node, the velocities of particles often adhere to a specific distribution, offering an indirect glimpse into their kinematic properties.
The essence of this methodology lies in establishing an error criterion for simulating scenarios that involve an infinite number of particles, thereby guiding the optimization of algorithms in a meaningful direction. As such, the omission of direct positional and velocity magnitude information is deemed acceptable within the confines of this algorithm.

Taking the battery as a case study, we can envision the electrons within it serving as particles for Molecular Dynamics simulations, subsequently transformed into a network architecture. With the positive and negative terminals aligned horizontally, the electrons positioned at varying lateral points encounter differing magnitudes of force. Consequently, we can quantify the average increase in their kinetic energy along the horizontal direction. Given that electrons inherently possess kinetic energy, we can leverage the principle of detailed balance to derive the transition probability between nodes.
It's noteworthy that as the lateral position shifts, the weights assigned to the nodes undergo corresponding changes. However, regardless of the lateral position, the total number of electrons migrating horizontally remains constant. This underscores the varying concentrations of electrons across different lateral regions.
In the realm of Molecular Dynamics, by seamlessly integrating the computations stemming from this algorithm with measurements of voltage and inter-terminal distance, we can establish a lower boundary for the inherent errors in simulating this process with molecular dynamics.

Assuming molecular dynamics to be highly efficient, could we possibly deduce microscopic information solely through the measurement of macroscopic parameters like temperature and voltage? The truth is, acquiring microscopic data is exceedingly difficult, which is why this paper adopts a network model to estimate such microscopic quantities.

Consider a battery configured with its anode and cathode horizontally aligned, where I assume a voltage $U$ and a distance $l$ separating the electrodes. This approach begins by estimating the average kinetic energy $E_{0}$ of electrons within the battery under conditions of zero voltage and low temperature. Subsequently, for points located at transverse distances $l_{1}$ and $l_{2}$ within the battery, the increment in kinetic energy can be expressed as: $U*l_{1}/l,U*l_{2}/l$.

At these $l_{1}$ and $l_{2}$ points, this work construct network models, enabling us to directly derive the weights of each network node through the partition function. Notably, the total number of electrons traversing the cross-section per unit time remains constant at both $l_{1}$ and $l_{2}$. The overall trend of electron transverse flow can be succinctly represented by $C_{2}+C_{3}-C_{1}-C_{4}$. Moreover, the average transverse velocities of electrons at $l_{1}$ and $l_{2}$ are designated as $v_{1}$ and $v_{2}$, respectively.

The detailed balance relationship inherent in the network structure depicted in the Fig.~\ref{fig:3}(a) allows us to derive the following equation.

\begin{equation}
\label{eq:6}
   v_{1} \frac{1-e^\frac{l_{1}}{l*T}}{1+e^\frac{l_{1}}{l*T}}=v_{2} \frac{1-e^\frac{l_{2}}{l*T}}{1+e^\frac{l_{2}}{l*T}}
\end{equation}

Let's contemplate a scenario in which, across various positions within the battery, the quantity of electrons that actually migrate comprises a minute portion of the total electron population at those locations. The vast majority of electrons harmoniously oscillate around their equilibrium points, whereas a small segment of these electrons, upon reaching the extreme right of their oscillation cycle, do not revert to their equilibrium positions but instead embark on a migratory journey.
Furthermore, if we posit that the transverse motion of electrons within the battery can be satisfactorily modeled using harmonic oscillators, then the ratio of the frequencies of these oscillators at $l_{1}$ and $l_{2}$ can be elegantly expressed as:

\begin{equation}
    V=\frac{1}{2}\mu w^2(\triangledown x)^2
\end{equation}
Here, $\mu$ denotes the mass of the harmonic oscillator, $w$ signifies the frequency, $\triangledown x$ represents the distance in the transverse direction, and $V$ stands for the potential energy.
This formulation sheds light on the frequency variations occurring at different locations, thereby enabling us to approximate the potential energy associated with electron interactions within the battery. By integrating the absolute values of the harmonic oscillator's velocity across various time points, we can derive the following relationship.
\begin{equation}
\label{eq:8}
    v_{1}=\frac{v_{max}}{w_{1}}=\frac{\sqrt{2(E_{0}+U*l_{1}/l)/\mu}}{w_{1}}
\end{equation}
\begin{equation}
\label{eq:9}
    \frac{w_{1}}{w_{2}}=\frac{v_{2}}{v_{1}} \frac{\sqrt{E_{0}+U*l_{1}/l}}{\sqrt{E_{0}+U*l_{2}/l}}
\end{equation}
Here, $v_{max}$ denotes the velocity at the equilibrium point, whereas $w_{1}$ and $w_{1}$ signify the frequencies corresponding to the locations $l_{1}$ and $l_{2}$, respectively.

To visualize the fluctuations in frequency, this paper adopts a simplified approach by assigning a value of 1 to macroscopic quantities, including the electron's mass and electric field intensity. Utilizing Eq.~\ref{eq:6} and Eq.~\ref{eq:9}, this work have constructed a trend graph, which is presented in Fig.~\ref{fig:3} (b) for clear illustration.

\section{Conclusion}

In parallel with the generalized Fourier functions, this paper presents a simulation algorithm that simulates an infinite number of particles, effectively transforming this vast system into a manageable, finite network of interconnected nodes. The simple interactions between individual particles are seamlessly translated into correlations among these network nodes. By leveraging the unique properties inherent in certain network structures or tailored network configurations, we can establish a rigorous lower bound for the simulation algorithm's error. This methodology offers a precise roadmap for enhancing and optimizing the simulation algorithm, while showcasing the broad applicability of this transformative approach.

In the context of the two-dimensional Ising model, an infinite lattice system is transformed into a finite set of 10 network nodes, with the interactions between lattice points intricately mapped onto complex correlations among these nodes. The utilization of Monte Carlo algorithms to model these intricate relationships between network nodes is both logical and efficient. Subsequently, by presenting a network structure that closely resembles the initial state, we are able to establish a rigorous lower bound for the error inherent in Monte Carlo simulations. When compared to the analytical solution of the two-dimensional Ising model, it becomes evident that the Monte Carlo simulation algorithm indeed exhibits a certain degree of error.

In the realm of Molecular Dynamics, particles' velocities and accelerations undergo dimensional decomposition. Considering a two-dimensional system as an illustrative case, particles are segmented into 4 distinct categories based on their velocity vectors, with each category corresponding to a unique network node. Additionally, the directionality of acceleration is similarly classified into 4 types, each associated with a different type of bond, thereby constructing the overall network structure. This framework also introduces potential metrics for evaluating the precision of Molecular Dynamics simulations. Lastly, the algorithm is further elaborated upon through a detailed case study involving batteries. Furthermore, it offers insights into estimating micro-level information from macro-level data. The crucial information of Molecular Simulation, namely the potential energy, has been quickly estimated.

\textbf{\textit{Acknowledgments---}}
\label{acknowledgments}
This paper is supported by the National Natural Science Foundation of China-China Academy of Engineering
Physics(CAEP)Joint Fund NSAF(No. U2230402).

\bibliography{sample}

\nocite{*}

\end{document}